\documentclass[amsmath,amssymb,pra,aps,mathbbm,superscriptaddress]{revtex4}
\usepackage{amsmath,amsfonts,amssymb,amsthm,graphics,graphicx,epsfig,bbm}
\usepackage[colorlinks=true,citecolor=blue,linkcolor=blue,urlcolor=blue]{hyperref}
\usepackage[usenames]{color}
\usepackage{graphicx}
\usepackage{subfigure}
\usepackage{epsfig}
\usepackage{bbold}
\usepackage{dcolumn}
\usepackage{bbm}
\usepackage{color}
\usepackage{verbatim}
\usepackage{epstopdf}
\usepackage{amssymb}
\usepackage{amstext}
\usepackage{latexsym}
\usepackage{hyperref}
\usepackage{amsfonts}
\usepackage{psfrag}
\usepackage{xcolor}
\usepackage{times}
\usepackage{mathtools}
\usepackage{braket}
\usepackage{soul}
\usepackage{stmaryrd}

\newcommand{\beq}{\begin{equation}}
\newcommand{\eeq}{\end{equation}}
\newcommand{\beqa}{\begin{eqnarray}}
\newcommand{\eeqa}{\end{eqnarray}}

\newcommand\numberthis{\addtocounter{equation}{1}\tag{\theequation}}

\begin{document}
\title{Dispersive Regimes of the Dicke Model}
\author{Diego Barberena}
\affiliation{Departamento de Ciencias, Secci\'on F\'isica, Pontificia Universidad Cat\'olica del Per\'u, Apartado 1761, Lima, Peru}
\author{Lucas Lamata\footnote{Corresponding author: lucas.lamata@gmail.com}}
\affiliation{Department of Physical Chemistry, University of the Basque Country UPV/EHU, Apartado 644, 48080 Bilbao, Spain}
\author{Enrique Solano}
\affiliation{Department of Physical Chemistry, University of the Basque Country UPV/EHU, Apartado 644, 48080 Bilbao, Spain}
\affiliation{IKERBASQUE, Basque Foundation for Science, Maria Diaz de Haro 3, 48013 Bilbao, Spain}
\date{\today}

\begin{abstract}
We study two dispersive regimes in the dynamics of $N$ two-level atoms interacting with a bosonic mode for long interaction times. Firstly, we analyze the dispersive multiqubit quantum Rabi model for the regime in which the qubit frequencies are equal and smaller than the mode frequency, and for values of the coupling strength similar or larger than the mode frequency, namely, the deep strong coupling regime. Secondly, we address an interaction that is dependent on the photon number, where the coupling strength is comparable to the geometric mean of the qubit and mode frequencies. We show that the associated dynamics is analytically tractable and provide useful frameworks with which to analyze the system behavior. In the deep strong coupling regime, we unveil the structure of unexpected resonances for specific values of the coupling, present for $N\ge2$, and in the photon-number-dependent regime we demonstrate that all the nontrivial dynamical behavior occurs in the atomic degrees of freedom for a given Fock state. We verify these assertions with numerical simulations of the qubit population and photon-statistic dynamics.
\end{abstract}


\maketitle

\section*{INTRODUCTION}
The Dicke model is one of the simplest models that captures the effects of the interaction between several atoms and an electromagnetic mode of radiation.  Since its original proposal \cite{Dicke}, it has been the subject of extensive investigations and was shown to possess a remarkable effect, namely, the existence of thermal~\cite{Hepp,Wang}, quantum~\cite{Emary} and excited state~\cite{Perez} quantum phase transitions at specific values of the temperature, coupling strength and excitation energy, respectively. Following those lines, there has been a plethora of studies discussing its spectral properties. This includes the cases of small $N$ \cite{Braak}, where integrability analyses have been done, and large but finite $N$~\cite{Chen}, with accompanying discussions about the relations between the aforementioned transitions and with chaos~\cite{Bastarrachea,Bastarrachea2,Bastarrachea3,Bastarrachea4}. The small coupling strength near the resonant regime has also been thoroughly investigated, resulting in the integrable Tavis-Cummings model \cite{Tavis} after performing a rotating-wave approximation. Several studies of the dynamics have been done with a variety of motivations, e.g., entanglement, collapses and revivals \cite{Alvermann, Agarwal}, behavior near the classical limit~\cite{Bakemeier}, and open system problems~\cite{Fuchs}. The dynamics in the small qubit frequency regime has also been explored \cite{Agarwal}, but mainly for low coupling strengths. In the experimental frontier, it has been implemented in a variety of scenarios \cite{Baumann1,Baumann2,Baden} and quantum simulations have also been considered, primarily for one qubit \cite{Ballester,Simon,Langford,Braumuller}, but also for larger $N$ \cite{Mezzacapo, Lamata}. Furthermore, current experimental trends are directioned towards reaching increasingly large values of the coupling \cite{Niemczyk,Pol, Yoshihara,Yoshihara2,Pol2}, such that it is worthwhile to analyze the large-coupling region and identify interesting effects. Recently, efforts to classify the quantum Rabi model in a variety of coupling regimes have been carried out~\cite{MikelRabi}.

In this Article, we study the dynamics of the Dicke model in the regime of small but nonzero qubit frequency $\omega_0$ for a finite number $N$ of atoms. The ensuing separation of timescales naturally leads to the use of an adiabatic approximation, a technique previously employed in the one qubit case \cite{Larson1, Irish} and for large qubit frequency \cite{Liberti,Relano} in the language of the Born-Oppenheimer approximation. This allows us to separate fast oscillating behavior at the frequency of the mode, $\omega$, from secular effects induced by the finiteness of $\omega_0$ and keep only first-order corrections. For a fixed and small value of the qubit frequency, there are two discernible regimes showing distinct behavior depending on the value of $g$. The first one is the deep strong coupling regime, introduced originally for the quantum Rabi model \cite{Casanova} and extended now to this multiqubit quantum Rabi model, which shows effects not present for the $N=1$ atomic case. The second one is an intermediate coupling regime where the atomic dynamics is nontrivial, governed by the Lipkin-Meshkov-Glick Hamiltonian \cite{Lipkin,Tsomokos}, yet the bosonic mode decouples for each number state. At the same time, both behaviors become mixed in the intermediate regime. 

\section*{RESULTS}

{\it{The model}.--} We consider now the Dicke Hamiltonian,
\begin{equation}
H=\omega_0 J_z+\omega a^{\dagger}a+gJ_x(a+a^{\dagger}),
\label{DickeHamiltonian}
\end{equation}
where $\omega_0$ is the qubit frequency, $\omega$ is the photon frequency, and $g$ is the coupling constant. $J_z$ and $J_x$ are collective atomic operators, defined as $2J_z=\sum_i \sigma_z^i$, $2J_x=\sum_i\sigma_x^i$. Both of them, together with $J_y$, satisfy the angular momentum commutation relations. We also introduce $\hat{J}^2$, whose eigenvalues are $J(J+1)$ and work in the symmetric subspace with $2J=N
$, where $N$ is the number of atoms. Likewise, $a$ and $a^{\dagger}$ are photon annihilation and creation operators, respectively.

In the regimes we are interested in, both rotating and counterrotating terms are important, and we must keep both to account for the dynamics. Instead of neglecting either of them, we perform a displacement transformation on $H$~\cite{Chen,Bastarrachea5}, and its associated time evolution operator,
\begin{align*}
U_H&=D(\lambda)U_{H'}D(-\lambda),\\
H'&=D(-\lambda)HD(\lambda).\numberthis
\end{align*}
By choosing $\lambda=\frac{g}{\omega}J_x$, such that it does not commute with $J_z$, the resulting $H'$ is
\begin{equation}
H'=\omega_0 e^{-\frac{gJ_x}{\omega}(a^{\dagger}-a)}J_z e^{\frac{gJ_x}{\omega}(a^{\dagger}-a)}+\omega a^{\dagger}a-\frac{g^2}{\omega}J_x^2.
\end{equation}
The reason for this transformation is that, if $\omega_0 = 0$, the Hamiltonian would be diagonal, as is apparent from the previous equation. If we introduce ladder operators with respect to $J_x$, $J_{\pm}=J_z\mp iJ_y$, and use the ensuing commutation relations, $J_xJ_{\pm}=J_{\pm}(J_x\pm 1)$, the Hamiltonian takes the simpler form,
\begin{equation}
H'=\frac{\omega_0}{2} \left(e^{\frac{g}{\omega}(a-a^{\dagger})}J_-+{\rm H.c.}\right)+\omega a^{\dagger}a-\frac{g^2}{\omega}J_x^2.
\end{equation}
Since $\omega\gg\omega_0$, we move to an interaction picture generated by the $\omega a^{\dagger}a$ term,
\begin{gather}
U_H=e^{\frac{gJ_x}{\omega}(a^{\dagger}-a)}e^{-i\omega t\, a^{\dagger}a }U_{H_2}e^{-\frac{gJ_x}{\omega}(a^{\dagger}-a)},
\raisetag{-.5em}\label{EvolOpt}
\end{gather}
where $U_{H_2}$ is the time evolution operator associated with the following Hamiltonian,
\begin{equation}
H_2=\frac{\omega_0}{2} \left[ e^{\frac{g}{\omega}(ae^{-i\omega t}-a^{\dagger}e^{i\omega t})}J_-+{\rm H.c.} \right] - \frac{g^2}{\omega}J_x^2.\label{Hamilt}
\end{equation}
Further manipulations depend on the magnitude of $\frac{g^2}{\omega^2}$. It should be noted that that the Hamiltonian of Eq.(\ref{Hamilt}) is structurally very similar to the trapped ion Hamiltonian before doing the vibrational rotating wave approximation. In consequence, it is amenable to the same methods of analysis~\cite{Vogel}, a fact we will exploit in the following sections.

{\it Deep Strong Coupling Regime.--}
$g\gtrapprox\omega$ in this scenario so $g^2/\omega$, the prefactor of the $J_x^2$ term in Eq.~(\ref{Hamilt}), is much larger than $\omega_0$, the prefactor of the other term. Thus, the former induces evolution on a faster time scale than the latter, but, at the same time, $g^2/\omega$ remains comparable to $\omega$ itself. Therefore, it is convenient to move to an interaction picture with respect to the $J_x^2$ term as well. The resulting time evolution operator and Hamiltonian are
\begin{eqnarray}
U_H=&&e^{\frac{gJ_x}{\omega}(a^{\dagger}-a)}e^{-i\left(\omega a^{\dagger}a-\frac{g^2J_x^2}{\omega}\right)t}U_{H_3}e^{-\frac{gJ_x}{\omega}(a^{\dagger}-a)},
\end{eqnarray}
\begin{equation}
H_3=\frac{\omega_0}{2} e^{\frac{g}{\omega}\left(ae^{-i\omega t}-a^{\dagger}e^{i\omega t}\right)}e^{i\frac{g^2}{\omega}(2J_x+1)t}J_-+{\rm H.c.}\label{h3Eq}
\end{equation}
We point out that, in Eq.~(\ref{h3Eq}), the resulting dynamics depends strongly on the magnitude $\frac{g^2}{\omega^2}$.

\begin{figure}
{\includegraphics[width=0.5\textwidth]{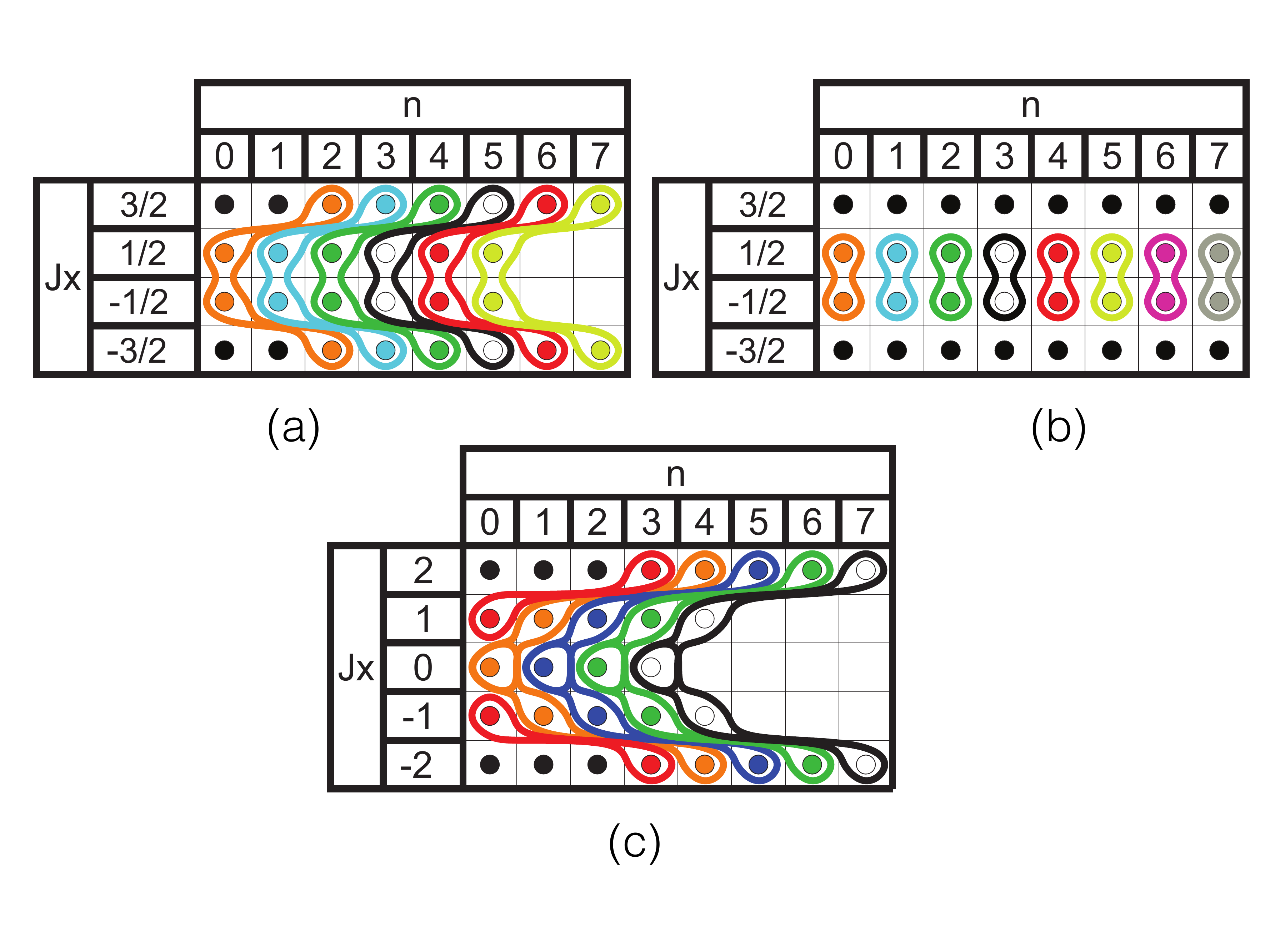}}
\caption{Schematic representation of the Hilbert space of the system. On each table, each square represents a basis state labeled by the eigenvalues of $J_x$ (left column) and of $a^{\dagger}a$ (top row). Solid lines indicate the chains of states that are coupled by the effective Hamiltonian for different values of $g$ and $J$: (a)  $g=\omega$, $J=\frac{3}{2}$, (b) $g$ non resonant, $J=\frac{3}{2}$, and (c) $g=\omega$, $J=2$.  Black dots represent decoupled states.}\label{Cadenas}
\end{figure}

\begin{figure}
\includegraphics[width=0.48\textwidth]{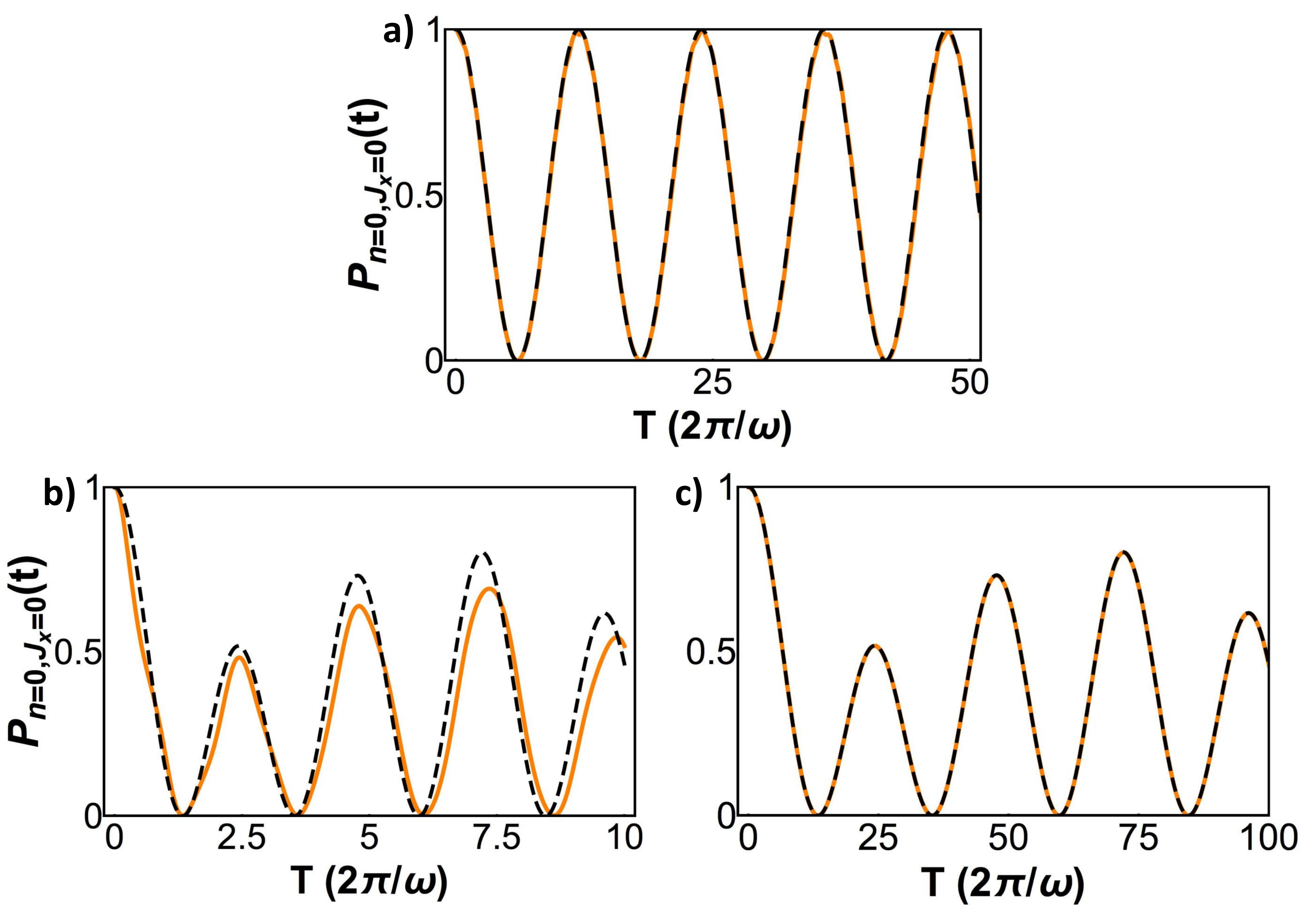}
\caption{Evolution of $P(t)$ for (a) $J=1$ (2 atoms), $g=\sqrt{5}\omega$, $\omega_0=0.1\omega$; (b) $J=4$ (8 atoms), $g=\omega$, $\omega_0=0.1\omega$ and (c) $J=4$ (8 atoms), $g=\omega$, $\omega_0=0.01\omega$ showing the existence of resonances. Numerical simulation is orange and theoretical approximation is dashed black.}\label{DSCResonance}
\end{figure}

{\it i) Resonant behavior.}
In the case of $g=\omega\sqrt{k}$ where $k$ is an integer, each $J_x$ eigenstate will generate a resonance with a different term in the displacement factor, as shown in the following table for the specific case of  $J=2$, namely, four qubits,
\begin{center}
\begin{tabular}{c||c|c|c|c|c}
    $J_x$ & $-2$  & $-1$ & $0$ & $1$ & $2$ \\ \hline\hline
    $(2J_x+1)k\omega$& $-3k\omega$ & $-k\omega$ & $k\omega$ & $3k\omega$& $5k\omega$
\end{tabular}
\end{center}
For example, a $3k\omega$ term will resonate with an $a^{3k}$ term. For this purpose, it is convenient to perform a Fourier expansion of the displacement operator,
\begin{align*}
e^{\beta(ae^{-i\omega t}-a^{\dagger}e^{i\omega t})}&=\Omega_{\hat{n}}^0(\beta)+\sum_{m>0}^{\infty}\Omega_{\hat{n}}^m(\beta)a^m e^{-im\omega t}+(-1)^me^{im\omega  t}a^{\dagger}{}^m\Omega_{\hat{n}}^m(\beta),\\[1.5em]
\Omega_{\hat{n}}^m(\beta)&=\beta^me^{-\frac{\beta^2}{2}}L_{\hat{n}}^m(\beta^2)\frac{\hat{n}!}{(\hat{n}+m)!},\numberthis
\end{align*}
where $L_n^m$ is an associated Laguerre polynomial. The first-order effective Hamiltonian is obtained by averaging with respect to $\omega$,
\begin{gather}
\begin{align*}
\braket{H_3}=\frac{\omega_0}{2} \biggl[ \Omega_{\hat{n}}^ka^k(P_{0}J_-P_{1})+\Omega_{\hat{n}}^{3k}a^{3k}(P_{1}J_-P_{2})-a^{\dagger}{}^{3k}\Omega_{\hat{n}}^{3k}(P_{-2}J_-P_{-1})-a^{\dagger}{}^{k}\Omega_{\hat{n}}^k(P_{-1}J_-P_{0})+{\rm H.c.}\biggr] ,
\end{align*}\numberthis
\raisetag{-.5em}
\end{gather}
where the $P_m$ are projectors onto the $J_x=m$ state. Note that the expected $a^{5k}$ term does not appear because $P_{2}J_-=0$. Nothing can be lowered to the $J_x=2$ eigenstate. Using states $\ket{m,n}$ for which $J_x=m$ and $a^{\dagger}a=n$, the dynamics can be seen to generate the following dispersive chain of connected states,
\begin{align*}
    \ket{-2,n+4k}\leftrightarrow& \ket{-1,n+k}\leftrightarrow \ket{0,n}\leftrightarrow\ket{1,n+k}\\&\leftrightarrow \ket{2,n+4k}.\numberthis
\end{align*}
The Hilbert space can thus be divided into subspaces as shown in Fig. \ref{Cadenas}(c).
For large $n$, these chains emerge and are connected as displayed in Fig.~\ref{Cadenas}. To verify these analytical results, we did numerical simulations of Eq.~(\ref{DickeHamiltonian}) for $N=2$, $g=\sqrt{5}\omega$ and $\omega_0=0.1\omega$, as shown in Fig.~\ref{DSCResonance}(a). We also computed the cases for $N=8$, $g=1$, $\omega_0=0.1\omega$ and $\omega_0=0.01\omega$, see Figs.~\ref{DSCResonance}(b) and~\ref{DSCResonance}(c). To achieve this, we plot
\begin{equation}
P(t)=|\bra{J_x=0,n=0}U(t)\ket{J_x=0,n=0}|^2.
\label{P(t)}
\end{equation}
We point out that, for $\omega_0=0.1\omega$, a complete depopulation of the initial state takes place in a short time. For smaller $\omega_0$, the approximation works better but the time span required to depopulate the initial state increases accordingly. In the $N=8$, $\omega_0=0.1\omega$ case, the approximation captures qualitatively the correct behavior, but higher-order secular effects become apparent as the theoretical approximation lags behind the simulations. Peaks are distorted as well due to micromotion effects.

This same resonant behavior is present for $J$ equal to a half-integer. In the specific case of $J=\frac{3}{2}$, the corresponding table reads
\begin{center}
\begin{tabular}{c||c|c|c|c}
    $J_x$ & $-3/2\hspace{0.1cm}$  & $-1/2\hspace{0.1cm}$ & $1/2$ & $3/2$ \\ \hline\hline
    $(2J_x+1)k\omega$& $-2k\omega$ & $0k\omega$ & $2k\omega$ & $4k\omega$
\end{tabular}
\end{center}
In this case, the $J_x=-\frac{1}{2}\leftrightarrow J_x=\frac{1}{2}$ transition involves no change in the photon number, as shown in Fig.~\ref{Cadenas}(a). When considering higher values of $J$, the dispersive chains start growing but remain finite.

{\it ii) Off-resonant behavior.}
When moving away from these resonances, the system responds differently depending on whether $J$ is an integer or a half-integer. If $J$ is an integer, the chain dynamics are progressively suppressed as the detuning increases and the system reaches a dispersive regime. The resonances described in the previous sections have a width of order $\omega_0$, such that they are sharper if $\omega_0$ is smaller. In the specific case of $J=1$, i.e., for two qubits, the minimum value of $P$ for a $k$ resonance can be easily calculated,
\begin{align*}
P_{\rm min}&=\frac{(g^2-k\omega^2)^2}{(g^2-k\omega^2)^2+4\omega^2\omega_0^2(\Omega_0^k)^2 k!}\\[1em]
&=\frac{(g^2-k\omega^2)^2}{(g^2-k\omega^2)^2+4\omega^2\omega_0^2(\frac{g}{\omega})^{2k}\frac{e^{-\frac{g^2}{\omega^2}}}{k!}} \, . \numberthis
\end{align*}
\\
It is Lorentzian in $g^2$, rather than $g$, such that the peak is not symmetric with respect to the resonance.
\begin{figure}
\includegraphics[width=0.48\textwidth]{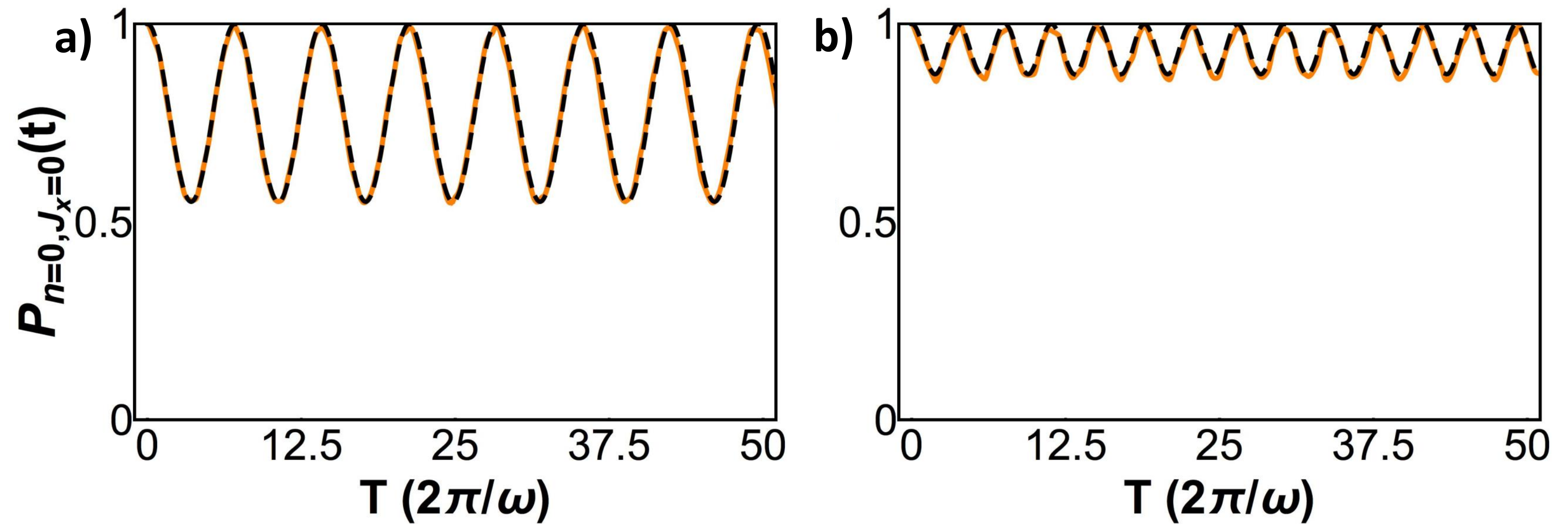}
\caption{Evolution of $P(t)$ for $J=2$ (4 atoms) and $\omega_0=0.1\omega$, with (a) $g=(\sqrt{3}+0.03)\omega$ and (b) $g=(\sqrt{3}+0.07)\omega$. Numerical simulation is shown in orange and theoretical approximation in dashed black.}\label{DSCOResonance}
\end{figure}
Figs.~\ref{DSCOResonance}(a) and~\ref{DSCOResonance}(b) show that as we move away from the resonance, the value of $P_{\rm min}$ departs from $0$ and starts to increase, while the oscillations experience a frequency shift. The new frequency for the $k^{th}$ resonance and $J=2$ reads
\begin{equation}
\frac{1}{2}\sqrt{(g^2-k\omega^2)^2+4\omega^2\omega_0^2\left(\frac{g}{\omega}\right)^{2k}\frac{e^{-\frac{g^2}{\omega^2}}}{k!}}.
\end{equation}
\begin{figure}
\includegraphics[width=0.48\textwidth]{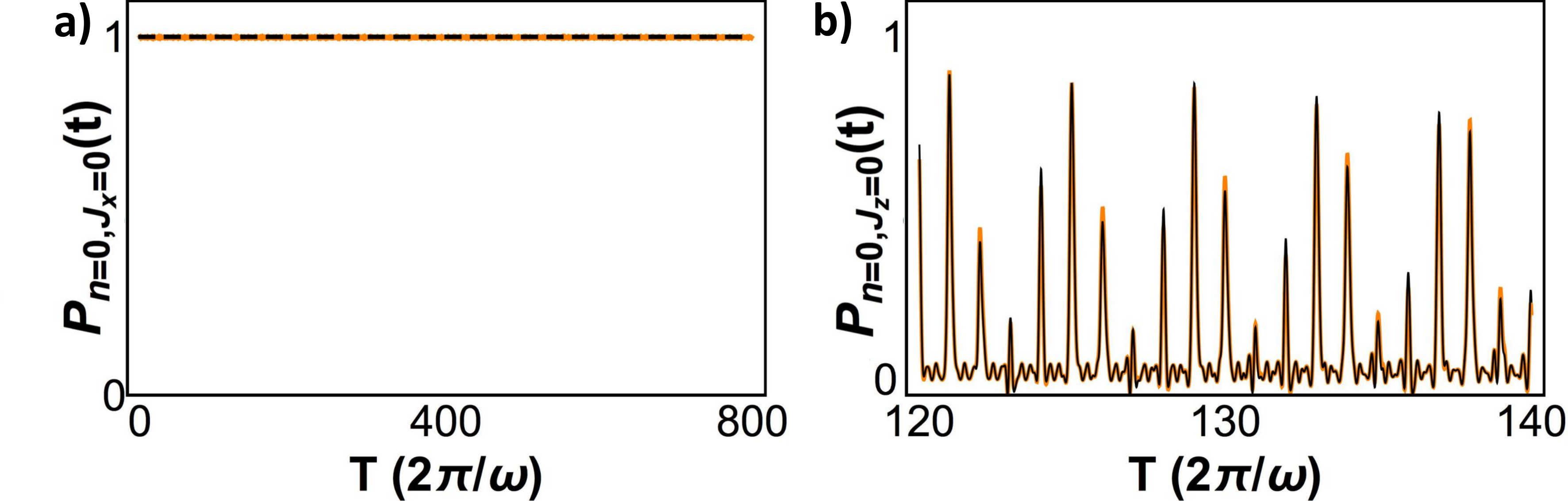}
\caption{ (a) Evolution of $P(t)$ for $g=1.2\omega$ and $\omega_0=0.01\omega$, for 800 cycles of the mode frequency. (b) Evolution of $|\bra{J_z=0,n=0}U(t)\ket{J_z=0,n=0}|^2$ for $g=1.2\omega$ with $\omega_0=0$ (black) and $\omega_0=0.01\omega$ (orange). The plot shows the evolution from $\omega t=(2\pi)120$ to $\omega t=(2\pi)140$.}\label{DSCOResonanceII}
\end{figure}
Figure~\ref{DSCOResonanceII}  shows that for a sufficiently detuned $g$, most of the unveiled dynamics is almost suppressed.

If $J$ is a half-integer, the physical description changes radically. As can be seen from the table for $J=3/2$, the $J_x=-1/2\leftrightarrow J_x=1/2$ transition is not suppressed by the rapidly oscillating terms, since $g^2(2J_x+1)/\omega=0$ for the $J_x= -1/2$ entry, independently of the value of $g$. The effective Hamiltonian in these conditions reads
\begin{equation}
\braket{H_3}=\frac{\omega_0}{2}\Omega_{\hat{n}}^0\left(P_{-\frac{1}{2}}J_-P_{\frac{1}{2}}+P_{\frac{1}{2}}J_+P_{-\frac{1}{2}}\right).
\end{equation}
This encompasses the case of the single-qubit quantum Rabi model ($J=1/2$) and accounts for the distorted peaks reported in Ref.~\cite{Casanova}. Graphically, the dispersive chains of Fig. \ref{Cadenas}(a) collapse into Fig. \ref{Cadenas}(b).

\begin{figure}
\includegraphics[width=0.48\textwidth]{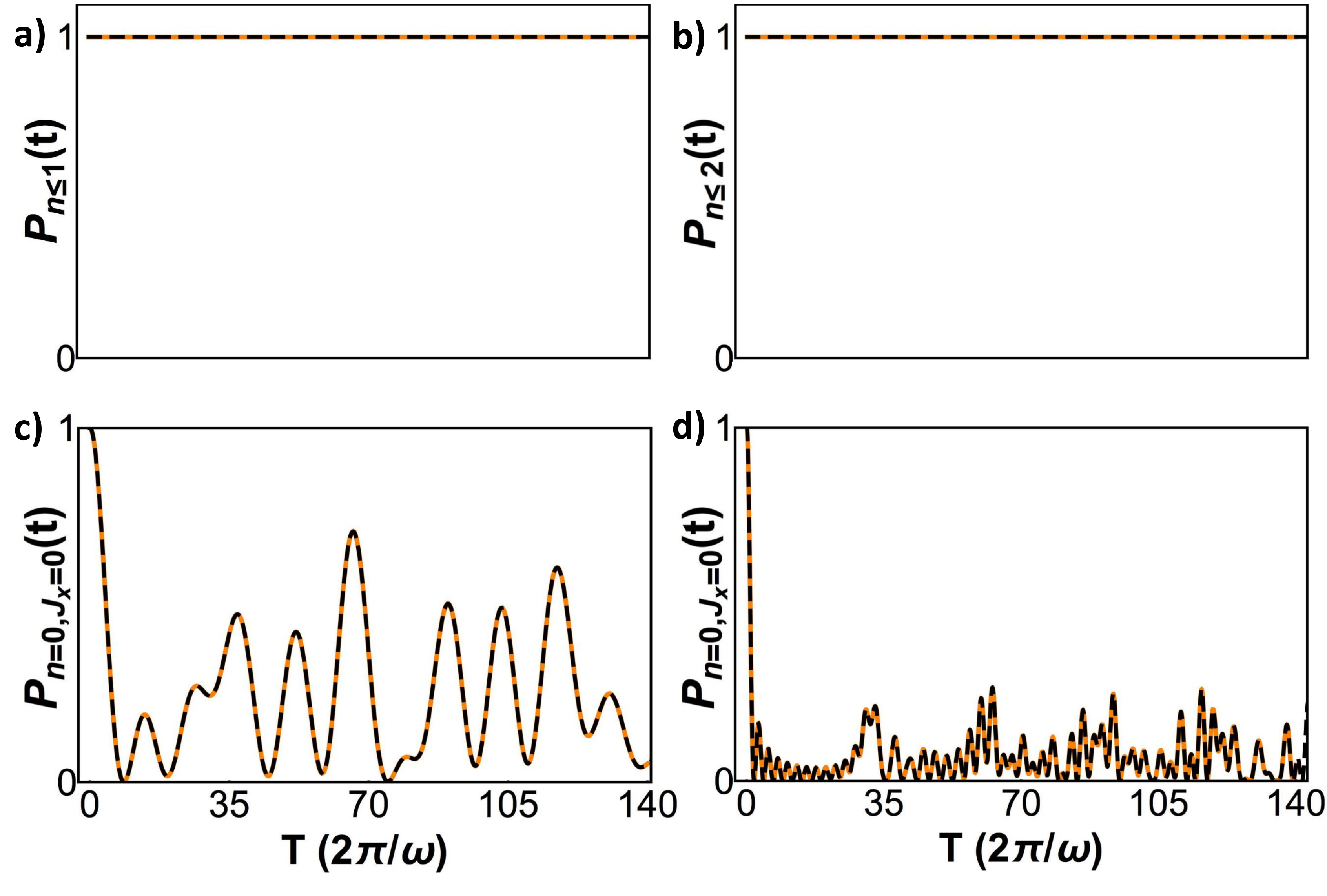}
\caption{For $g^2J=0.01\omega^2$ and $\omega_0=0.01\omega$, we show (a) the probability of finding one or less photons at time $t$ for $J=4$ (8 atoms), (b) the probability of finding two or less photons at time $t$ for $J=20$ (40 atoms), (c) $P(t)$ for $J=4$, (d) $P(t)$ for $J=20$. Orange curves are numerical simulations and theoretical approximation is plotted in dashed black.}\label{Atomic}
\end{figure}

{\it Photon-number-dependent Regime.--}
A different behavior arises when $g\ll\omega$, such that $g^2\approx\omega\omega_0$. Then, both terms in Eq.~(\ref{Hamilt}) are comparable, and it is not useful to go into the interaction picture generated by the $J_x^2$ term. Instead, the first-order effective Hamiltonian is obtained by averaging directly,
\begin{equation}
\braket{H_2}=\omega_0\Omega_{\hat{n}}^0J_z-\frac{g^2}{\omega}J_x^2\label{EffHamilt2},
\end{equation}
 and the result, as has been pointed out before \cite{Larson2, Morrison, Unanyan}, is a Lipkin-Meshkov-Glick Hamiltonian. Within this approximation, Fock states remain unchanged, while each Fock state induces a different evolution on the atomic internal states, characterised  by the $\Omega_{\hat{n}}^0$ coefficient accompanying $\omega_0$. The evolution of the Hamiltonian of Eq.~(\ref{DickeHamiltonian}) changes the number of photons due to the displacement operators in Eq.~(\ref{EvolOpt}), albeit very weakly since $\frac{g}{\omega}$ is now a small quantity. This is verified in Figs. \ref{Atomic}(a) and \ref{Atomic}(b). Furthermore, the effective Hamiltonian is a good approximation even for $40$ atoms as shown in Figs. \ref{Atomic}(c) and \ref{Atomic}(d), while still displaying nontrivial physics. In a way, the dispersive chains of the previous section have become vertical in the tables of Fig. \ref{Cadenas}. If the initial state contains several Fock components, the $\Omega_{\hat{n}}^0$ term originates the dispersive dynamics that gives rise to the collapses and revivals reported in Ref.~\cite{Agarwal}.
It should be noted that this regime is close to the critical coupling of the Dicke quantum phase transition. To analyse this, it is convenient to discuss thoroughly the approximations that led into Eq.~(\ref{EffHamilt2}). The separation of timescales that justifies the averaging requires that the resulting effective dynamics does not occur at a rate comparable to the fast timescale. The fastest effective dynamics occur at frequencies of the order of $2J\omega_0$ and $\frac{g}{\omega}^2J^2$, since they are at first glance the largest energy differences that occur in $\braket{H_2}$. Thus, reasonable upper bounds are $2\omega_0J\lessapprox0.1\omega$ and $\frac{g}{\omega}^2J^2\lessapprox0.1\omega$.

As an example, let us consider now 100 atoms. The upper bounds for $\omega_0$ and $g$ are then $10^{-3}\omega$ and $3\times 10^{-3}\omega$, respectively. If we take $\omega_0=10^{-6}\omega$, then the critical coupling can be calculated,
 \begin{equation}
 g_{\rm crit}=\sqrt{\frac{\omega_0\omega}{N}}=10^{-4}\omega,
 \end{equation}
 which is within the approximation.

\section*{DISCUSSION}

In this work, we have developed a useful framework to analyze the dispersive dynamics of the low-qubit frequency region of the Dicke model. This can be phrased in terms of effective Hamiltonians that generate dynamics in isolated dispersive chains of the system Hilbert space, whose detailed structure depends on the value of the coupling.

Several remarks are in order. (i) The quantity plotted in most graphs, P(t), defined in Eq.~(\ref{P(t)}), is particulary convenient for showing the possible discrepancies between the complete and the effective evolutions because the state $\ket{J_x=0,n=0}$ is invariant under the displacement operators of Eq.~(\ref{EvolOpt}). However, a general initial state will be affected by such transformations and the complete evolution is bound to be complicated because the displacement parameter depends on a state's value of $J_x$. This would mask the simple nature of the effective dynamics. (ii) While the Lipkin-Meshkov-Glick model resulting from the adiabatic elimination of the photon field is well known, the results of this paper consider it as a special case of the dispersive chain dynamics in which transitions among spin states are not accompanied by changes in photon number. Thus, effects that have been discussed in this setting may be investigated in the context of more complex chains. (iii) Corrections to the effective dynamics described in the paper are of two kinds: there are secular corrections that appear as additional contributions to the effective Hamiltonians and there are small (of order $\frac{\omega_0}{\omega}$) periodic (with frequency $\omega$) coherent transitions to other states.

 \section*{AUTHOR CONTRIBUTIONS}
D.B developed this work and wrote the manuscript. L.L. suggested the seminal ideas and contributed to the analysis. E.S. helped to check the feasibility and improved the ideas and results shown in the paper. All authors have carefully proofread the manuscript. E.S. supervised the project throughout all stages.

\section*{ACKNOWLEDGMENTS}
The authors acknowledge support from Spanish MINECO/FEDER FIS2015-69983-P, Ram\'on y Cajal Grant RYC-2012-11391, and Basque Government IT986-16.

\section*{ADDITIONAL INFORMATION}
The authors declare no competing financial interests.

\end{document}